# Developing a cyber security culture: Current practices and future needs

Betsy Uchendu [1], Jason R.C. Nurse [1], Maria Bada [2] and Steven Furnell [3]

[1] *University of Kent, UK*
[2] *University of Cambridge, UK*
[3] *University of Nottingham, UK*

Corresponding author: Jason R.C. Nurse, j.r.c.nurse@kent.ac.uk

**Highlights**

- A systematic review of the last ten years of research on security culture, information security culture and cyber security culture
- Top management support, security policy, awareness/training key to building cyber security culture
- Developing a security culture requires in-depth knowledge of organisation and employees
- Questionnaires and surveys often used to measure culture, but have several weaknesses which should not be overlooked
- Very few cyber security culture frameworks/approaches have been evaluated in the real-world

**Abstract**

While the creation of a strong security culture has been researched and discussed for decades, it continues to elude many businesses. Part of the challenge faced is distilling pertinent, recent academic findings and research into useful guidance. In this article, we aim to tackle this issue by conducting a state-of-the-art study into organisational cyber security culture research. This work investigates four questions, including how cyber security culture is defined, what factors are essential to building and maintaining such a culture, the frameworks proposed to cultivate a security culture and the metrics suggested to assess it. Through the application of the PRISMA systematic literature review technique, we identify and analyse 58 research articles from the last 10 years (2010-2020). Our findings demonstrate that while there have been notable changes in the use of terms (e.g., information security culture and cyber security culture), many of the most influential factors across papers are similar. Top management support, policy and procedures, and awareness for instance, are critical in engendering cyber security culture. Many of the frameworks reviewed revealed common foundations, with organisational culture playing a substantial role in crafting appropriate cyber security culture models. Questionnaires and surveys are the most used tool to measure cyber security culture, but there are also concerns as to whether more dynamic measures are needed. For practitioners, this article highlights factors and models essential to the creation and management of a robust security culture. For research, we produce an up-to-date characterisation of the field and also define open issues deserving of further attention such as the role of change management processes and national culture in an enterprise's cyber security culture.

**Keywords**: Cybersecurity culture, information security culture, security awareness, organisational culture, management, business, SMEs, behaviour, psychology.

# 1. Introduction

Over the last decade, the topic of security culture has featured heavily in practice and research as organisations look to combat an increase in attacks that exploit human factors. Within this domain, several terms have been used including security culture, information security culture and more recently, cyber security culture. Although cyber security is often used synonymously with information security more generally, the nature of the two terms arguably differs. For example, Von Solms and Van Niekerk (2013) define cyber security as the aim to protect an additional set of assets, particularly human and organisational assets, which can be regarded as more extensive. It therefore may be seen to consider shared values, beliefs and expected behaviours regarding the protection of this wide set of aspects (Ioannou et al., 2019; Da Veiga, 2016a).

In this paper, we contribute to the practitioner and academic communities by critically investigating the state-of-the-art of cyber security culture research through a systematic review of ten years of related literature. This work seeks to summarise key research developments within an area that still proves challenging for businesses in their pursuit of building robust security cultures (ComputerWeekly, 2019; Tripwire, 2020). Specifically, we apply a rigorous systematic review process with the aim of examining how the use of terminology has changed, the factors central to building a cyber security culture, and the frameworks and tools used to cultivate and measure it. This research is novel as compared to other related security culture reviews (e.g., Glaspie & Karwowski, 2017; Nasir et al., 2019a; Sas et al., 2020) through its combined coverage and analysis of these three areas (instead of disparate studies), meta-analysis of pertinent, recent research (to understand the origins and context of articles published and provide more general insight into the direction of the field), and identification of current outstanding issues as we enter a new decade. We also differentiate our work from other articles focused specifically on security policy compliance (e.g., Bulgurcu, Cavusoglu & Benbasat, 2010; Moody, Siponen & Pahnila, 2018; Cram, D'arcy & Proudfoot, 2019; Paananen, Lapke & Siponen, 2020) as this study aims to be wider in scope, with policies and compliance discussions primarily considered through articles that examine security culture. This is a similar approach to existing culture studies (Nasir et al., 2019a; da Veiga et al., 2020), and is beneficial as it guides our scope and contributions.

The remainder of this article is structured as follows. Section 2 outlines and explains the PRISMA protocol used to conduct the review. In Section 3, the results of the article selection process are presented as well as a review of the articles – both their details and contributions as it pertains to our research aim. Next, Section 4 engages in an in-depth analysis of the results discovered, reflects on open issues in security culture research, outlines practical implications, and considers our findings in the context of related prior reviews. Section 5 concludes the article and summarises our insights for research and practice.

# 2. Methodology

## 2.1 Protocol and registration

The present research used the Preferred Reporting Items for Systematic reviews and Meta-Analysis (PRISMA) guidelines (Moher et al. 2009). PRISMA outlines a robust protocol which has demonstrated use in security research (Knight and Nurse, 2020). Such systematic review techniques use carefully designed methods to select and review relevant research literature and analyse the results that emerge from the research. The aim is to synthesise understanding using a valid and reliable technique (Okoli and Schabram, 2010). Our work draws on PRISMA's checklist, as well as a four-phase flow diagram, to help guide the analysis. The checklist defines the main sections and topics that need to be presented and addressed in the review, while the flow diagram includes protocols related to identification of sources and articles, screening of articles, determining eligibility, and inclusion of relevant articles.

## 2.2 Eligibility criteria

This systematic review used the following eligibility criteria to help decide which research articles were relevant. This inclusion criteria (IC) included:

- IC1, which required only articles published in English.
- IC2, if IC1 was met, involved articles pertaining to and extensively covering the broader topic of cyber security culture (inclusive of related terms such as security culture and information security culture).
- IC3, if the article had met the first two criteria, included articles that were specific to organisations; i.e. organisations in general or users in organisations, often employees. This allowed our research to exclude articles that concentrated on national or the public's security culture.

## 2.3 Information sources

The search was applied to ACM Digital Library, IEEE Xplore, ScienceDirect, Web of Science, Scopus and ProQuest libraries. ACM Digital Library, IEEE Xplore, Science Direct and Web of Science cover the widest body of research on computer science and more specifically cyber security. Scopus and ProQuest were useful databases in capturing research articles from non-computing domains. These additional libraries helped increase the chances that all relevant articles were considered; organisational culture, for instance, touches on a wider scope of topics such as psychology and business management. The choice of these six article repositories and publisher databases was preferred as compared to other options (such as Google Scholar, for instance), due to the higher chances of finding relevant, peer-reviewed articles. Our search was run in August 2020 and covered articles published from January 2010 to August 2020. This period would capture the most recent research as well as allowing us to consider the emergence of cyber security culture as compared to more dated terms including information/IT security culture.

## 2.4 Study selection

Using the four phases outlined in the PRISMA diagram, the study selection process followed these steps:

1. Electronic databases were searched, focusing on literature that pertained to cyber security in organisations. The keyword query was designed to use terms that related to the topic of interest and would allow the widest capture of articles. Specifically, the following search was used across the databases: ("security culture" OR "cybersecurity culture").
2. Duplicates both within and across databases were then removed following the collection of results. The title, abstract and keyword listing of each research article was reviewed, and articles were filtered manually and marked as appropriate if inclusion criteria were met. If not, the relevant criteria were noted to record exclusion reason.
3. The remaining articles that were not excluded in step 2 were either partially or completely reviewed and further refined based on the inclusion criteria. Once the relevant articles were determined, these were further narrowed down to only include articles from 2010 and onwards, limiting the review period to the last 10 years.

4. Reference lists of relevant literature were scanned to include any studies that may have been missed. These were subject to the same actions in steps 2 and 3, and added if the inclusion criteria and the publication year requirement was met.

These actions were performed and reviewed amongst authors. This allowed any disparities to be discussed and agreed.

## 2.5 Data items collected

The review focused on extracting data from articles that were associated with examining how the use of security culture terminology has changed, the factors central to building a cyber security culture, and the frameworks and tools used to cultivate and measure it. These are further detailed in Table 1.

**Table 1. Data items**

| Data Item | Description |
|---|---|
| Research foundation | <ul><li>Definitions and characteristics of cyber security culture (inclusive of security culture and information security culture)</li><li>Factors that are regarded as important to build and maintain a cyber security culture</li><li>Approaches/tools/frameworks used or proposed to build a cyber security culture</li><li>Metrics used or proposed to assess cyber security culture</li></ul> |
| Key findings and contributions | <ul><li>Key findings and contributions of the research into cyber security culture</li></ul> |
| Study/experiment details | <ul><li>Research instruments used to conduct the study/experiment, if any (e.g., interviews, surveys, theory-based research)</li><li>Type of participants and number of participants involved (e.g., employees, top-level management, IT departments)</li><li>Sector or industry of focus (e.g., finance, healthcare)</li></ul> |
| Author(s) and publication venue details | <ul><li>Country author(s) is/are affiliated with</li><li>Type of publication (conference or journal)</li><li>Name of publication venue (i.e., title of conference or journal)</li><li>Year of publication</li></ul> |

# 3. Results

## 3.1 Study selection

In total, 58 research articles fulfilled the criteria outlined and thus were selected for review. The keyword search across the databases resulted in 1931 articles. From these, 536 were removed as they were duplicates. Following this, the first inclusion criteria (IC1) was applied, and 30 articles were discarded. After assessing the title, abstracts and keywords of the remaining articles, 1110 articles were excluded as they did not meet the criteria (IC2) of referring to cyber security culture (which, as discussed in Section 1, we view as inclusive of information security culture). The full text of the remaining 255 articles were reviewed against the final criteria (IC3), and a further 200 results did not meet the criteria and thus were excluded. Three additional articles fit the criteria but were excluded because we were unable to access the full text. The reference lists of the final 52 articles were examined, and after checking the inclusion criteria, a further 6 papers were added to the final set. Figure 1 illustrates a PRISMA flow diagram outlining the process used to achieve the final 58 articles.

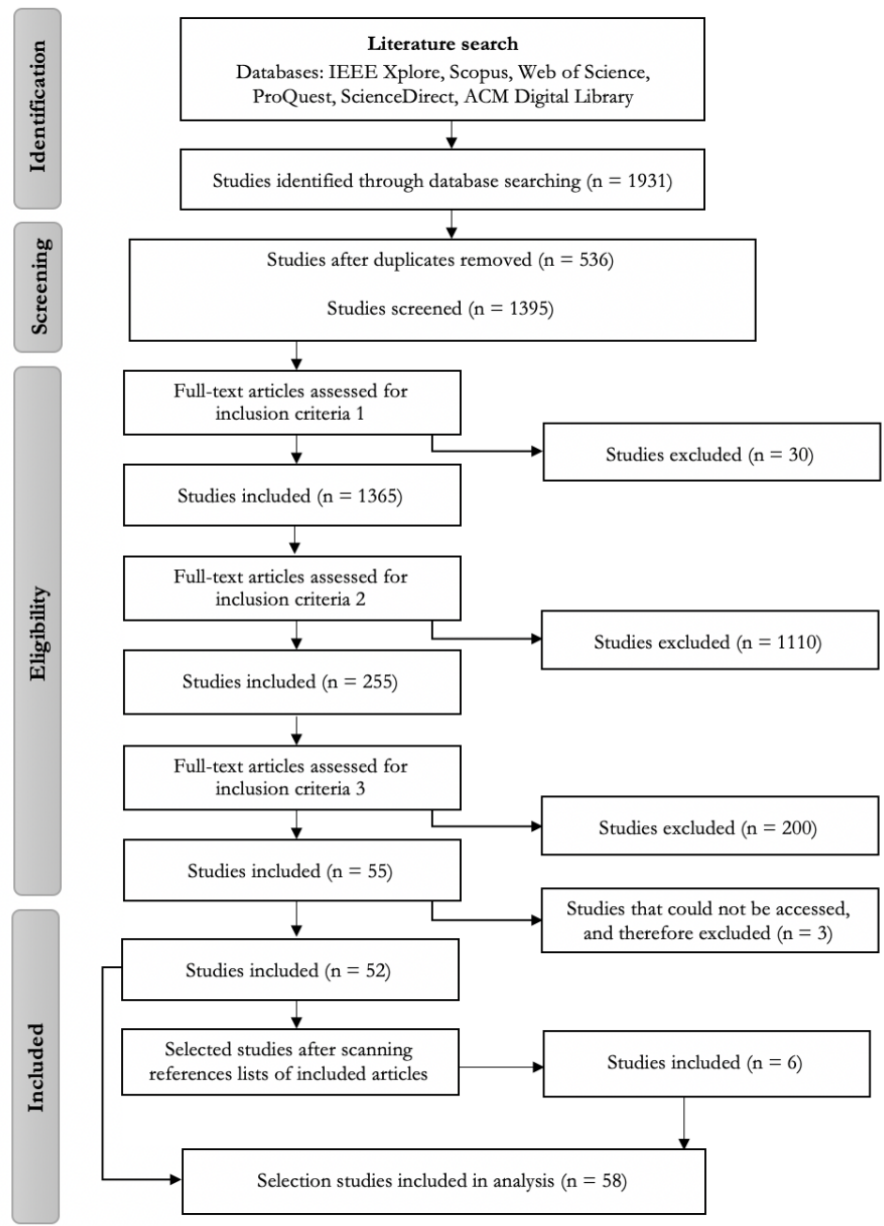

**Figure 1.** PRISMA flow diagram presenting the inclusion process

## 3.2 Study characteristics

This section of the study summarises the results from the meta-analysis on the final set of articles. It considers the distribution of papers across conferences and journals, and the prevalence of certain publication venues, countries (of authors), research instruments and other article details. Firstly, we present the full set of research articles in Table 2, and use paper identification numbers (IDs) (e.g., P1, P2) throughout our work to allow easier reference of review articles.

**Table 2. Full list of research articles reviewed and respective paper IDs**

| Paper ID | Study | Paper ID | Study |
|----------|-------|----------|-------|
| **P1** | Alfawaz et al. (2010) | **P30** | Da Veiga (2016a) |
| **P2** | Da Veiga and Eloff (2010) | **P31** | Da Veiga (2016b) |
| **P3** | Lacey (2010) | **P32** | Hassan and Ismail (2016) |
| **P4** | Lim et al. (2010) | **P33** | Santos-Olmo et al. (2016) |
| **P5** | Sánchez et al. (2010) | **P34** | Tang et al. (2016) |
| **P6** | Van Niekerk and Von Solms (2010) | **P35** | Da Veiga and Martins (2017) |
| **P7** | Batteau (2011) | **P36** | Gcaza et al. (2017) |
| **P8** | Alnatheer et al. (2012) | **P37** | Hassan et al. (2017) |
| **P9** | Hassan and Ismail (2012) | **P38** | Masrek et al. (2017) |
| **P10** | Olivos (2012) | **P39** | Da Veiga (2018) |
| **P11** | Shahibi et al. (2012) | **P40** | Masrek et al. (2018a) |
| **P12** | AlHogail and Mirza (2014a) | **P41** | Masrek et al. (2018b) |
| **P13** | AlHogail and Mirza (2014b) | **P42** | Mokwetli and Zuva (2018) |
| **P14** | Astakhova (2014) | **P43** | Nævestad et al. (2018) |
| **P15** | D'Arcy and Greene (2014) | **P44** | Ioannou et al. (2019) |
| **P16** | Da Veiga and Martins (2014) | **P45** | Marotta and Pearlson (2019) |
| **P17** | Lopes and Oliveira (2014) | **P46** | Nasir et al. (2019b) |
| **P18** | Reid et al. (2014) | **P47** | Nel and Drevin (2019) |
| **P19** | Reid and Van Niekerk (2014) | **P48** | Pătraşcu (2019) |
| **P20** | AlHogail (2015a) | **P49** | Ruhwanya and Ophoff (2019) |
| **P21** | AlHogail (2015b) | **P50** | Tolah et al. (2019) |
| **P22** | Alnatheer (2015) | **P51** | Van't Wout (2019) |
| **P23** | AlKalbani et al. (2015) | **P52** | Alshaikh (2020) |
| **P24** | Da Veiga (2015) | **P53** | Blythe et al. (2020) |
| **P25** | Da Veiga and Martins (2015) | **P54** | Da Veiga et al. (2020) |
| **P26** | Greig et al. (2015) | **P55** | Govender et al. (2020) |
| **P27** | Lim et al. (2015) | **P56** | Nasir et al. (2020) |
| **P28** | Martins and Da Veiga (2015) | **P57** | Schneider et al. (2020) |
| **P29** | Sherif et al. (2015) | **P58** | Wiley et al. (2020) |

**Paper distribution.** Six out of 58 articles (~11%) were published in 2010, one in 2011 (~2%), and four in 2012 (~7%). No articles were included from 2013. Eight out of 58 (~14%) were published in 2014, and the highest number of articles included were from 2015 at ten (~17%). Five articles (~8%) were included from 2016, four from 2017 (7%), five from 2018 (~8%) and eight from 2019 (~14%). At the time of our search in August 2020, seven articles (~12%) were included for review.

**Publication type.** Figure 2 displays a simplified visual of the number of studies and their publication type (journal or conference) in each year. Of the 58 articles, 32 papers were published at conferences, and 26 published in journals.

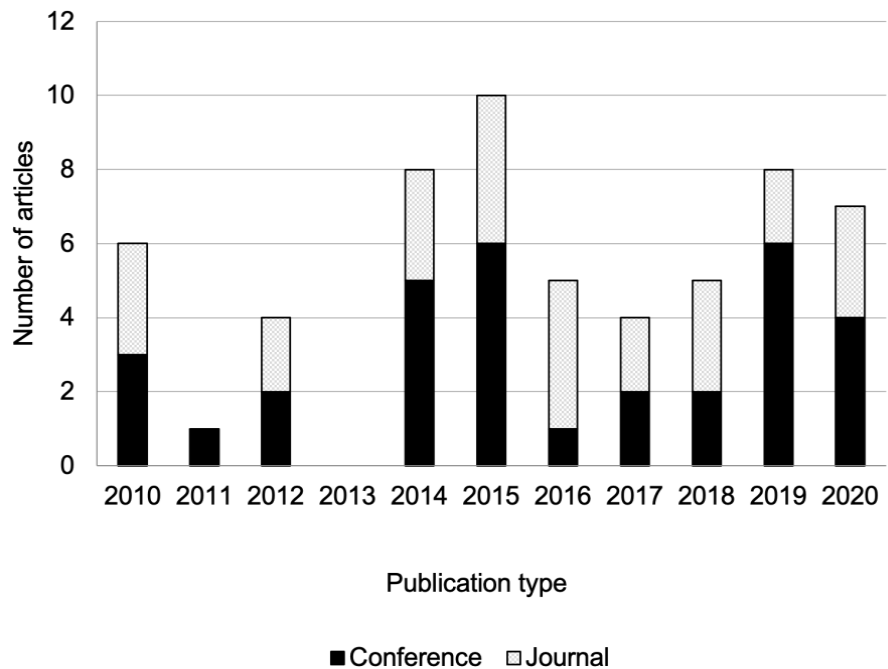

**Figure 2.** Years articles published and publication type

**Publication venue.** Table 3 was created to summarise the number of papers by source of publication venue. A range of journals and conferences, including Computers & Security, Pacific Asia Conference on Information Systems (PACIS), Journal of Theoretical and Applied Information Technology (JATIT), International Symposium on Human Aspects of Information Security and Assurance (HAISA), International Information Security for South Africa Conference, Information Management & Computer Security, and Americas Conference on Information Systems (AMCIS), had two or more articles within this review. The two most frequently appearing journals, Computers & Security, and Information and Computer Security, publish research in the field of security. Computers & Security issue articles aimed at IT security professionals to provide expertise and practical advice on management of systems. Information and Computer Security focuses on organisational and business domains of information security. The ethos of both journals is consistent with culture research and is likely the reason for the high total of published papers from this review.

**Table 3. Number of selected studies, by publisher**

| Source | Total | Studies |
|---|---|---|
| Computers & Security | 7 | P2, P6, P25, P35, P52, P54, P58 |
| Information and Computer Security (formerly Information Management & Computer Security) | 6 | P3, P15, P31, P36, P39, P47 |
| Pacific Asia Conference on Information Systems (PACIS) | 3 | P4, P8, P23 |
| Journal of Theoretical and Applied Information Technology (JATIT) | 3 | P11, P13, P32 |
| International Symposium on Human Aspects of | 3 | P10, P24, P28 |
| Information Security and Assurance (HAISA) | | |
| International Information Security for South Africa Conference | 2 | P18, P19 |
| Americas Conference on Information Systems (AMCIS) | 2 | P45, P57 |
| International Conference on ENTERprise Information Systems | 1 | P5 |
| International Journal of Business Anthropology | 1 | P7 |
| International Conference on Information, Communication Technology and System (ICTS) | 1 | P12 |
| Australasian Conference on Information Security (AISC) | 1 | P1 |
| European Conference on Information Management and Evaluation (ECIME) | 1 | P16 |
| World Conference on Information Systems and Technologies (WorldCIST) | 1 | P17 |
| Scientific and Technical Information Processing | 1 | P14 |
| International Conference on Information Technology - New Generations (ITNG) | 1 | P22 |
| Computers in Human Behavior (CHB) | 1 | P20 |
| International Conference on Human Aspects of Information Security, Privacy and Trust (HAS) | 1 | P29 |
| International Journal of Security and its Applications | 1 | P21 |
| International Journal of Cyber Warfare and Terrorism (IJCWT) | 1 | P27 |
| Procedia - Social and Behavioral Sciences | 1 | P9 |
| Future Internet | 1 | P33 |
| Science and Information Computing Conference (SAI) | 1 | P30 |
| International Conference on Research and Innovation in Information Systems (ICRIIS) | 1 | P37 |
| International Conference on Education and Social Sciences (INTCESS) | 1 | P38 |
| International Conference on Advances in Big Data, Computing and Data Communication Systems (icABCD) | 1 | P42 |
| International Journal of Civil Engineering and Technology | 1 | P40 |
| International Journal of Mechanical Engineering and Technology | 1 | P41 |
| International European Safety and Reliability Conference (ESREL) | 1 | P43 |
| Information Security Journal: A Global Perspective | 1 | P46 |
| International Conference on Social Implications of Computers in Developing Countries | 1 | P49 |

| International Conference on Cyber Security and Protection of Digital Services | 1 | P44 |
|---|---|---|
| IFIP World Conference on Information Security Education | 1 | P50 |
| International Conference on Cyber Warfare and Security (ICCWS) | 1 | P51 |
| International Scientific Conference eLearning and Software for Education | 1 | P48 |
| World Congress on Internet Security (WorldCIS) | 1 | P26 |
| Information Technology and Management | 1 | P34 |
| International Conference on Information Management (ICIM) | 1 | P56 |
| Computer Science On-line Conference (CSOC) | 1 | P55 |
| International Conference on HCI for Cybersecurity, Privacy and Trust | 1 | P53 |

**Author distribution.** Figure 3 represents the geographical distribution of the author(s); where authors are affiliated with different countries, all locations have been represented. 22 studies were carried out by authors located in South Africa, over a third of the research in this review. From reviewing the data, it is clear that there are certain prolific researchers in that country that have found ways to work with organisations and industry to push forward security culture research and practice. The remainder of the authors are located across Asia-Pacific, Europe, North and South America and the Middle East. It is worth noting the relatively low number of articles from the US (3) and Europe (e.g., UK, France, Spain, Portugal and Norway have a total of 12) who are typically significant contributors to cyber security research.

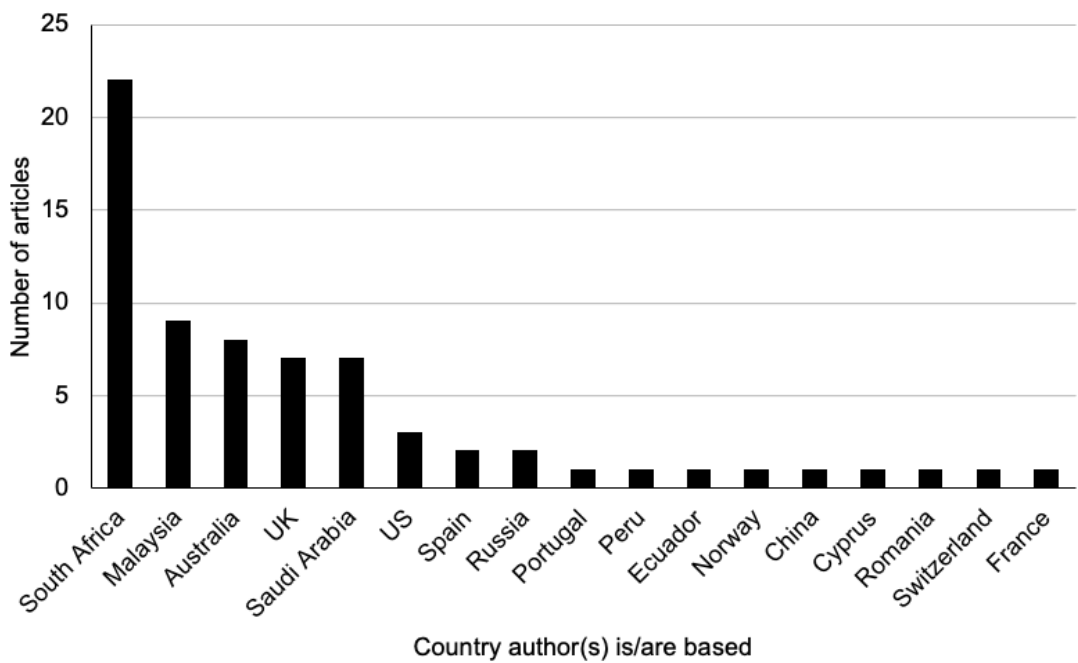

**Figure 3.** Geographical distribution of research authors

**Research instruments.** Table 4 presents the research instruments used across the 58 papers. As shown, a range of instruments were found spanning questionnaires and surveys, interviews, and theory-based research (e.g., literature reviews); this is further discussed in Section 3.6.

**Table 4. Research instruments used within the studies**

| Research instruments | Studies |
|---|---|
| Interviews | P1, P4, P10, P15, P21, P26, P27, P32, P34, P30, P37, P43, P50, P57 |
| Questionnaires and Surveys | P2, P3, P8, P11, P15, P16, P17, P20, P21, P23, P24, P25, P26, P27, P28, P30, P31, P34, P35, P39, P40, P41, P42, P44, P46, P47, P53, P54, P56, P58 |
| Theory-based research | P1, P2, P3, P4, P6, P7, P8, P9, P11, P10, P18, P12, P13, P17, P14, P19, P20, P21, P22, P27, P28, P29, P30, P33, P35, P36, P38, P42, P40, P41, P45, P46, P48, P49, P50, P51, P52, P54 |
| Other(s) | P5, P10, P33, P35, P45, P49, P55 |

**Participants.** Across the studies, most articles that incorporated employee participation used a range of participants from top management to employees. One article solely included business owners and managers (P42). Looking particularly at IT employees, eight articles included IT specialists, experts or information security managers as participants (P1; P8; P17; P20; P21; P44) or their departments (P10; P50). Two articles involved Human Resource (HR) managers (P4; P27).

**Sector or industry of focus.** There was no sector or industry predominant across the review. Where mentioned, case studies within papers included a wide variety of organisations such as manufacturing, higher education institutions (HEIs), and finance and insurance firms. Concentrating on industries, there were three articles focused solely on the healthcare industry (P9; P32; P37), two on banking and finance (P4; P45), one on retail (P26), and six on public organisations, i.e., governments and HEIs (P4; P21; P23; P36; P40; P46). The other selected papers did not concentrate on a specific industry.

**Security culture domain.** Security culture domain explores the differing terms surrounding cyber security culture and their respective applications. Figure 4 displays the proportion of articles that exclusively discuss security culture, information security culture or cyber security culture. The majority examine information security culture (42 in total), in comparison to only ten studies that consider cyber security culture. Seven studies analyse security culture, with no direct reference to information or cyber security. Cyber security culture articles have increased in the last 5 years, with a rise in articles in 2019. Where a study reflects on multiple domains, the paper has been represented in all appropriate categories.

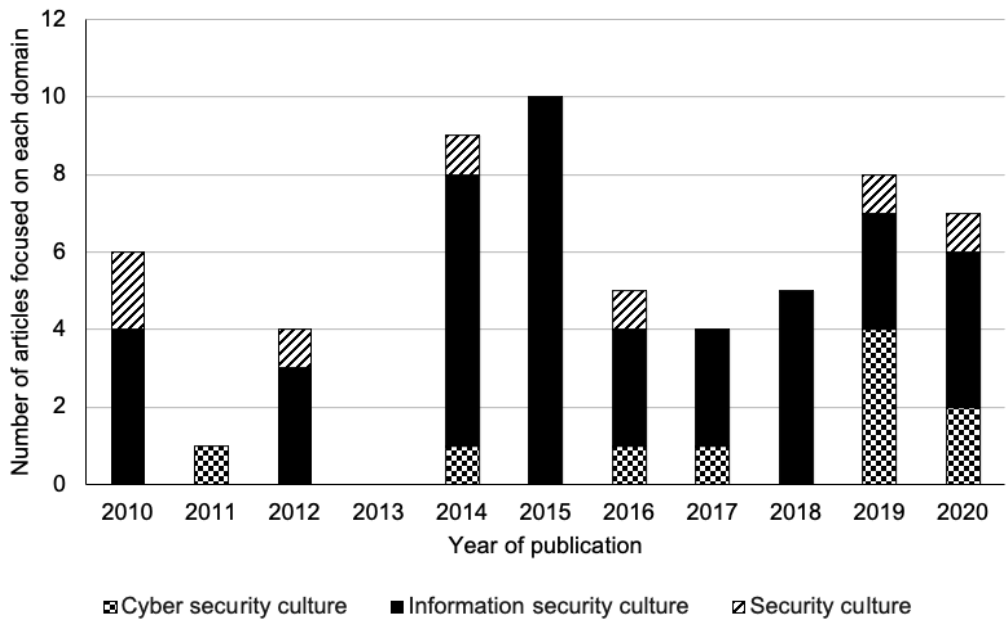

**Figure 4.** Distribution of articles that discuss each security culture term and the year of publication

A majority of the articles state a definition of cyber security culture by which their research aligns within the study. Additionally, not all studies aim to provide a method in which security culture can be measured or assessed. Most articles do present a framework or approach in which a security culture can be built and maintained. A majority (42 papers) focus on developing frameworks and tools, and analysing approaches to build an information security culture. A total of ten articles (P9; P11; P15; P22; P23; P29; P31; P34; P47, P58) present either multiple variables, or present findings focusing on one factor that influences cyber security culture. When exploring cultural metrics, 32 articles include a measuring instrument or tool to assess security culture. However, only three articles are focused primarily on defining security culture within their work. These points are discussed in detail in the next section.

## 3.3 Definitions and characteristics of cyber security culture

In total, five of the articles reviewed (P14; P35; P52; P54; P58) were predominantly focused on defining, characterising, or conceptualising cyber security culture (all-encompassing information security culture and security culture). To present these, this work focuses on definitions of information security culture, security culture and cyber security culture emerging from the research.

Most of the papers (49 studies) had a general definition of security culture or described its characteristics. Although not included in this review due to their publication date, two early definitions of information security culture appear to have guided definitions throughout the last ten years. The first definition by Dhillon (1999), described it as, "the totality of human attributes such as behaviours, attitudes and values that contribute to the protection of all kinds of information in a given organisation", supporting the idea that human characteristics play a part in information security culture. The second definition by Martins and Eloff (2002), regards it as an "assumption about perceptions and attitudes that are accepted in order to incorporate information security characteristics as the way in which things are done in an organisation, with the aim of protecting information assets", which further alludes to the role employee behaviour plays in the safeguarding of data.

Exploring information security culture definitions further, values appeared in 20 papers, such as P17, P20, and P42. Protection also features in 14 papers; e.g., P5, P33, P54. The attitudes of employees were referred to by 13 papers, including P10, P11 and P27. Assumptions appeared in the definitions of 13 papers, for example P9, P35 and P58. Within 12 papers, beliefs surrounding the characterisation of information security culture are mentioned, such as P15, P23, P34. The perception of employees and their behaviour appeared in eight papers, including P21, P38, and P54.

It is notable that only seven papers (P7; P14; P30; P36; P44; P45; P48) specifically define cyber security culture. Some cyber security culture definitions are similar to information security culture definitions, such as those featured by P44 and P46. Ioannou et al. (P44) describes cyber security culture as, "procedures laid down by an organization to all its employees, directing their course of action in all situations related to data integrity" and Nasir et al. (2019b) (P46) defines information security culture as emphasising, "the security of information assets by improving employees' information security behavior". Both suggest encouraging improved employee security behaviour particularly in relation to data/information.

In other cases, however, cyber security culture definitions tend to be broader in scope, and are less specific to the protection of information. Da Veiga (2016a) (P30), for example, characterised cyber security culture as something that, "promotes or inhibits the safety, security, privacy, and civil liberties of individuals, organizations or governments", which covers wider ground than solely the security of data, but also includes keeping people and organisations secure in their entirety. The same applies to the definition provided by Alshaikh (2020) (P52) who characterised cyber security culture as "contextualized to the behavior of humans in an organizational context to protect information processed by the organization through compliance with the information security policy and an understanding of how to implement requirements in a cautious and attentive manner as embedded through regular communication, awareness, training and education initiatives".

Astakhova (2014) (P14) posits that information security culture and cyber security culture are not synonymous. This is because the core of information security culture is its protection of security and data/information. However, the two concepts are intertwined and can be seen as separate parts of a whole.

Several studies reviewed include a definition with a specific focus on organisations. Eight papers (P2; P8; P43; P47; P49; P52; P54; P58) incorporate organisational culture within the definition of security culture. P8, P47 and P58 describe security culture as a subculture of organisational culture. The latter characterise information security culture as encompassing socio-cultural measures such that information security becomes a natural part of an employee's daily activities. Mokwetli and Zuva (2018) (P42) characterise the concept as interrelating with corporate systems and procedures. These definitions all call attention to the organisational foundation that cyber security culture needs to be built upon.

Another common theme in security culture definitions is security awareness, often viewed as an aspect of security culture (P3; P47; P54). Security awareness is described as the conditions in which users within an organisation are both committed to and aware of the security mission (Siponen, 2000). The interaction between security awareness and security culture appears often in information security research and these factors will be picked up on in the next section.

## 3.4 Cyber security culture factors

Overall, ten papers (P9; P11; P15; P22; P23; P29; P31; P34; P47; P58) explicitly studied the factors, or a particular factor, deemed as critical to building or maintaining a good, or positive, security culture. Others included factors within their discussions. Table 5 displays the most frequently occurring and pertinent factors appearing in the literature, and the related studies for further reference. We also include general definitions and descriptions for each factor listed in the Appendix; this assists in explaining how these terms are used.

**Table 5. Factors most often regarded as important to build and maintain a cyber security culture**

| Factors | Total | Studies |
|---|---|---|
| Top management support, leadership or involvement | 34 | P1, P2, P3, P4, P7, P8, P12, P13, P15, P16, P20, P21, P22, P23, P28, P24, P25, P29, P27, P30, P35, P37, P38, P39, P40, P41, P42, P44, P46, P47, P49, P50, P54, P57 |
| Security policy | 27 | P1, P2, P3, P4, P8, P10, P16, P22, P24, P25, P27, P28, P29, P30, P31, P32, P34, P35, P37, P38, P39, P40, P41, P44, P47, P50, P54 |
| Security awareness | 24 | P1, P4, P8, P9, P22, P23, P24, P25, P27, P28, P29, P30, P31, P32, P35, P37, P38, P39, P40, P41, P46, P47, P50, P54 |
| Security training | 21 | P1, P4, P7, P8, P10, P13, P17, P22, P28, P25, P24, P27, P30, P35, P38, P39, P43, P46, P47, P50, P54 |
| Change management | 12 | P2, P9, P13, P25, P24, P30, P35, P39, P45, P47, P51, P54 |
| Compliance | 12 | P10, P17, P22, P28, P34, P38, P40, P41, P46, P47, P50, P54 |
| Knowledge | 11 | P6, P7, P9, P10, P12, P33, P37, P35, P46, P51, P54 |
| Accountability and responsibility | 9 | P12, P20, P21, P23, P34, P38, P45, P47, P48 |
| Security risk | 8 | P22, P27, P35, P46, P47, P49, P50, P54 |
| Commitment | 8 | P1, P15, P17, P23, P37, P38, P43, P46 |
| Communication | 8 | P7, P12, P13, P15, P27, P34, P47, P44 |
| User management | 7 | P17, P19, P24, P25, P29, P30, P39 |
| Motivation | 7 | P12, P15, P17, P27, P38, P46, P47 |
| Trust | 6 | P7, P25, P30, P39, P47, P54 |
| National culture | 5 | P12, P35, P36, P49, P54 |
| Ethical conduct | 4 | P12, P22, P47, P50 |
| Regulations | 4 | P12, P20, P21, P42 |

| Establishing a network of champions | 1 | P52 |
|---|---|---|
| Rewards and sanctions | 1 | P53 |

The most discussed factor, particularly in terms of maintaining a positive security culture, is the support and leadership from top management. Thirty-four studies mentioned the importance of management, or leadership expectations to ensure security culture is at the forefront of an organisation and its culture. Notable examples of such studies include P1, P8, P12, P14, P23, P27, P38, P47, P54 and P57. For instance, Masrek et al. (2017) (P38) includes management support as a factor which is subdivided into two dimensions; information security importance and information security commitment, exemplifying the need for full support and involvement from top management.

A further 27 studies highlighted the importance of having clear policies and procedures for employees to understand and abide by. P27, P31, P37, P38, P44, and P47 are examples of papers that demonstrate the need for clear and well-communicated policy. Employee understanding often relates to policies and procedures; knowledge of these requirements is a vital part of both building and maintaining security culture, which is mentioned by 11 studies such as P7, P10, P12, P37 and P51. Security awareness and training programs are often studied together in research. A total of 24 studies (including P1; P38; P46; P54) in some way consider security awareness and its effect on security culture, and 21 studies (such as P1; P22; P27; P50) highlight that training is essential to increase security awareness and thus to cultivate culture. Wiley et al. (2020) (P58) studied the relationship between information security awareness and training; their findings support a positive relationship between both security training and security awareness, and also security awareness and security culture. These findings generally give evidence to support the often-assumed link between these two concepts.

One of the more intriguing factors, mentioned by 12 studies, is the management of change. AlHogail and Mirza (2014b) (P13) discussed that although many studies propose factors important to security culture, few have mentioned the transition process towards cultivating information security culture within an organisation. Arguably, previous studies have not provided concrete advice as to how to make changes if security culture is not at the positive level needed to protect information and systems. Change management is seen by Nel and Drevin (2019) (P47) and Van't Wout (2019) (P51) as a critical factor for inciting actual change needed to build and maintain security culture.

Along with understanding policies and procedures, compliance is a key aspect for employees in ensuring the values of an organisation are maintained; a total of 12 studies include compliance as a factor that influences security culture. For instance, P10, P17, P38, and P54 all discuss employees actioning and complying with policies. Compliance here focused on the process of ensuring employees and the organisation as a whole adheres to standards and regulations on security. Other highly discussed factors are accountability and responsibility (P12; P20; P23; P34; P38; P45), and user behaviour or management (P17; P19; P24; P25; P29; P30). Reid and Van Niekerk (2014) (P19) introduce the role of the user, dividing this concept into the nature of the task, the behaviour of the user and the psychology of the user. As these factors are frequently mentioned across multiple papers, this demonstrates the importance of employee responsibility to their behaviour within an organisation. This also lines up with the definitions of security culture covered in Section 3.3.

Commitment appears in eight papers (P1; P15; P17; P23; P37; P38; P43; P46) and generally pertains to the activities needed to ensure 'buy in' to allow policies to be adhered to and a strong culture fostered. It is apparent that trust, acknowledged by six papers (P7; P25; P30; P39; P47), is also seen as a factor important to building security culture. Trust is discussed in many ways across articles. P7 links trust to the confidence in employee actions and intentions. In addition, P25 relates to the protection of the organisation's information from a human perspective, the safekeeping of private information and trust in the communications of the organisation. Moreover, in some articles, trust is not clearly defined (e.g., P30) and which makes it difficult to broadly appreciate who is trusting whom (e.g., employee trusting the organisation or vice versa) and what the trust specifically relates to. This highlights one of the challenges of drawing conclusive insights from the existing body of work. Alongside trust, commitment across employees suggests that the relationship between an organisation, its management and its users can determine the state of the security culture.

In order to accomplish behaviour change and therefore establish a cyber security culture, Blythe et al. (2020) (P53) discuss the need for rewards and sanctions as part of a cyber security awareness campaign. Rewards or sanctions are practiced to bridge the knowledge-practice gap in cyber security awareness. This may also be considered as part of the motivation point above, depending on how it is applied. Another factor that has been identified as important in developing cyber security culture is that of establishing a network of champions (Alshaikh, 2020). As suggested, the cyber security champion role is to support the team in amplifying the security awareness messages, helping employees to adopt security behaviours identified from the policies as well as identifying SETA needs (skills, knowledge, and behaviours) and finally reporting progress back to the security team.

## 3.5 Cyber security culture approaches, tools and/or frameworks

A total of 37 studies explored approaches, tools, or frameworks regarding building or maintaining a security culture. P11, P38 and P50 focus on theoretical frameworks to guide the development of such security cultures. Conceptual information security culture models were developed by P9, P28, P29, P42, and P40. Although these are predominantly theoretical, they provide a basis of understanding for organisations to understand how a positive security culture can be cultivated and strengthened. For instance, Masrek et al. (2017) (P38) proposes a conceptual framework that decomposes such cultures into six dimensions, including management support, policy and procedure, compliance, awareness, budget and technology. This is an example of a framework built of dimensions that can be divided into factors that affect culture.

Schein's corporate culture model (1999) played a significant role in models developed by Van Niekerk and Von Solms (2010) (P6) and Reid and Van Niekerk (2014) (P19). The basis of this model is that there are three levels of corporate culture; artifacts, espoused values and shared tacit assumptions. P6 proposed that to create information security culture, an additional level of information security knowledge should be added to support the other three levels, as knowledge is a key part of a security culture. Therefore, according to this model, corporate culture is the basis of the Information Security Culture Model. Both studies use Schein's corporate culture model with the additional level of knowledge as a basis of understanding the structure of security culture; (P6) suggest that a good security culture would have each of the four levels higher than the "minimum acceptable baseline".

The use of the STOPE framework by Bakry (2004) was also proposed by three studies to act as the foundation, or "building blocks" upon which a framework could be built to guide the development of a security culture (P12; P20; P21). AlHogail (2015b) (P21) combined the Information Security Culture Framework from AlHogail (2015a) (P20) and the STOPE framework to create a new approach to understanding information security culture. The combined framework consists of five dimensions, which are Strategy, Technology, Organization, People, and Environment, and the aim is to cover four domains of human behaviour factors; preparedness, responsibility, management, and society and regulations (AlHogail and Mirza, 2014a). It also includes change management principles to encourage and support the needed change within organisations.

Change management processes and models have been proposed by P18, P39 and P51. Change management relates to persuading employees to change their behaviour to better align with what the organisation requires (Van't Wout, 2019). Da Veiga (2018) (P39) explores a holistic approach to change management with the Information Security Culture Change Management (ISCCM) approach. This aims to move organisations towards a positive security culture by managing the risks presented by human factors in the safeguarding of information. It is constructed of four phases to manage and implement behaviour change. Additionally, the Information Security Culture Change Framework by AlHogail and Mirza (2014a) (P12) focuses on cultivating information security culture through modifying behaviour, by integrating change management principles and examining human elements.

The ARCS framework by Govender et al. (2020) (P55) suggests that the key drivers to understanding, managing, supporting and changing information security culture are driven by understanding the current state of information security, investing in specific initiatives that help to reduce cost while also reducing information security risk, and creating a sustainable approach to managing and improving information security culture. The framework uses a set of questions on the quality and efficacy of the state of information security assessments, cost reduction and culture within the organisation.

Within the papers reviewed, a small subset (P5; P33; P42; P49) target Small and Medium Enterprises (SMEs) and Small, Medium and Micro Enterprises (SMMEs). SMEs are organisations that often struggle with limited resources, therefore tools developed for larger firms are not always usable due to affordability or complexity (de Araújo Lima, Crema and Verbano, 2020). Supported in the research, Santos-Olmo et al. (2016) (P33) explain that many frameworks created to improve information security culture are oriented towards larger organisations and lack factors necessary for smaller enterprises. Mokwetli and Zuva (2018) (P42) proposed a conceptual information communication and technology (ICT) security culture model made up of three categories; organisational, environmental and technological. The aforementioned study suggests that the adoption of the ICT model helps to minimise human error in SMMEs.

While there are no articles that appear to outwardly state that the approach has been developed solely towards large organisations, studies such as Lacey (2010) (P3) and Tang et al. (2016) (P34) do specify that the projects undertaken have been carried out in larger organisations. Da Veiga et al. (2020) (P54) states small, medium and large organisations are represented in the study's sample. Lacey (2010) (P3) reiterates that even large organisations often have low budgets and resources to direct towards developing effective change campaigns, which would be even more so the case for SMEs. More generally however, this does raise questions regarding the applicability of approaches for enterprises of varying sizes.

# 3.6 Cyber security culture metrics

A total of 36 studies attempted to assess cyber security culture. Examples of these include P26, P30, P40, P45 and P49. Questionnaires and surveys were the most frequently used tool to allow for some level of measurement. Lacey (2010) (P3) provides some reasoning behind the use of questionnaires, explaining that these instruments can provide useful metrics, allowing organisations to measure understanding and consider which areas of employee knowledge need intervention.

Across the review, 17 studies used individual questionnaires or surveys to assess employees within organisations, primarily concentrating on their knowledge of policy. Pertinent studies span P8, P15, P44, P57 and P53. Questionnaire items can be constructed in numerous ways; D'Arcy and Greene (2014) (P15) utilised previously validated items from existing measures, while Alnatheer et al. (2012) (P8) extracted questionnaire items utilising both expert feedback and responses from interviews to develop the instrument. Notable examples of questionnaire items in the latter include, "It is my responsibility to protect the information of my organisation" and "I am always educated or trained about new security policies". Most studies, including the three examples mentioned above, used a five-point Likert scale to measure responses; 1 as strongly disagree and 5 as strongly agree. One study (P23) used a seven-point Likert scale.

The Information Security Culture Assessment (ISCA) instrument developed by Da Veiga and Eloff (2010) (P2) aims to both assess and monitor information security culture within an organisation. It is constructed from the Information Security Culture Framework (Da Veiga and Eloff, 2010) with seven information security component categories; these include leadership and governance, security management and operations, security policies, security program management, user security management, technology protection and operations, and change. The instrument consists of 85 statements, such as, "I understand how information security is managed in ABC to protect information" and "ABC is committed to information security in order to protect information" (Da Veiga and Eloff, 2010), all of which were measured on a five-point Likert scale. This demonstrates the type of questions that assess awareness and perception regarding protection of information.

Da Veiga and Martins (2015) (P25) also used the ISCA questionnaire instrument, relying on items which focused on the human aspects of information security culture. An example item is, "I believe I have a responsibility regarding the protection of ABC's information assets (e.g., information and computer resources)", thus substantiating that a strong cyber security culture is considered as when employees are protecting and governing data at all times. The assessment focuses on understanding where employee knowledge, behaviour and attitudes lie; this is what is perceived to shape the security culture.

One instrument proposed by Da Veiga (2016a) (P30) is the cyber security culture research methodology (CseCRM). At the time of their study, a reliable tool was needed to measure cyber security culture and understand the level of risk due to human aspects. In this methodology, one of the key stages is the cyber security culture measuring instrument. The questionnaire used consisted of 11 dimensions, some of which included change, privacy perception, user management and cybersecurity in practice. Two example items cover, "I know what the risk is when opening emails from unknown senders, especially if there is an attachment" and "I use my work e-mail address on social networking sites", measured on a five-point Likert scale.

Other types of tools were also used to assess security culture; for example, short tests can be seen in two studies (P5; P33). These were developed for SMEs with the aim to help create and cultivate a positive security culture within smaller companies, without leading to high costs to maintain over time. Santos-Olmo et al. (2016) (P33) used a testing process which included security-related questionnaires. The user had to answer 50% or more of the questions correctly for the level of knowledge to be deemed as adequate. A 50% threshold in itself, being a potentially debatable approach to judgement.

Training sessions in combination with surveys are also used to assess information security culture, as seen in Olivos (2012) (P10). To assess knowledge, the study used a questionnaire to examine each individual's learning style to ensure each type was taken into account during the session. They proceeded to utilise training groups which involved employees taking a pre-test, engaging in a presentation, and taking a post-test to test for new knowledge. For instance, at the beginning of the session, only three out of ten employees agreed that they, "would not provide their password to another person under any circumstance". After the presentation, all participants knew not to provide their password.

Table 6 documents the various types of metrics and some of the respective assessment instrument used to assess cyber security culture generally. We note here that in some cases assessment types (e.g., specific questionnaires) have been reused by some authors, such as P31 and P35.

**Table 6. Metric area of focus, assessment type and the relevant articles**

| Metric focus generally | Assessment type(s) | Studies |
|---|---|---|
| **Awareness** | Questionnaires | P31, P35, P39, P58 |
| **Behaviour** | Questionnaires | P2, P3, P53 |
| **Knowledge** | Questionnaires | P16, P21, P24, P25, P28, P30, P43 |
| | Tests | P5, P33 |
| **Organisational/ environmental dimensions** | Process/Questionnaires | P2, P35, P58 |
| **Organisational performance** | Process/Observation/ Management tools | P45 |
| **Knowledge/behaviour** | Questionnaire/Observation/ Interview | P26 |
| | Training session | P10 |
| | Interviews | P1, P57 |
| | Questionnaire/Roundtable | P49 |
| **Security culture constructs/factors generally** | Questionnaires/Surveys | P2, P8, P11, P15, P17, P20, P23, P31, P34, P35, P40, P41, P42, P44, P46, P47, P54, P56, P58 |

# 3.7 Key findings and contributions of prior research

Upon reviewing the selected studies, the majority aim to contribute or extend on existing knowledge in the field. A recent study by Da Veiga et al. (2020) (P54) aimed to consolidate the existing perspectives on information security culture to try and inform a single definition of the topic. The paper states that it is a guide to future work; it is cognisant of the issue of no universally agreed upon definition and the need to bring together approaches, frameworks and measurement tools present up to this point to standardise information security culture.

For other studies (P4; P7; P23; P40), highlighting the importance of top management leadership and support was critical; the number of papers that mention this as a factor of security culture demonstrates this. Lim et al. (2010) (P4) concluded as with most articles that implementing and managing a healthy information security culture is difficult without the engaged involvement of top management.

In addition, Batteau (2011) (P7) proposed that without creating a community and environment of trust, upon which cyber security culture can be built effectively, it is difficult to implement and cultivate it. Ioannou et al. (2019) (P44) also identifies the importance of building a feeling of trust, not only with management and the organisation, but also among personnel within an employee's team. Organisational aspects are also focused on by Sherif et al. (2015) (P29) and Lopes and Oliveira (2014) (P17), who suggest that all aspects, from organisational, human and even to social, need to be considered when it comes to maintaining a cyber security culture. Key findings were also found in change management by three studies (P13; P39; P50). AlHogail and Mirza (2014b) (P13) findings suggest there is a crucial need for organisations to incorporate change management principles if a secure information security culture is to be built, due to the fact that actively seeing the correct behaviours in employees is a key part of positive change.

Marotta and Pearlson (2019) (P45) concluded that it is vital to influence employees and cultivate a "solid and effective human firewall" to foster a strong cyber security culture. Lacey et al. (2010) claims that the field must move away from typical security culture procedures and processes and towards observing and engaging with users. Van't Wout (2019) (P51) extends this idea, proposing a holistic approach towards developing and maintaining an organisational cyber security culture to design, develop and implement effective training and education programmes to ensure an improved cyber security culture. Another study (P58) found a positive relationship between organisational culture, security culture and security awareness. These findings all support the need for a broader approach.

Another noteworthy area within this review was SMEs. Four studies (P5; P33; P42; P49) focused on these types of companies and how to assess their security culture. These act to acknowledge that SMEs may be different and require more tailored approaches. To add to this, it was intriguing that the discussion on rewards and punishments is presented only by one recent study (P53). The study identified that even if some kind of rewards (e.g., gifts and public recognition) or punishments (e.g., disciplinary warning, restriction of privileged access) are provided as part of an awareness campaign, they might not be used or applied by employees. Therefore, an effective culture campaign might not be effective with just the provision of these as incentives.

# 4. Discussion

The purpose of this research is to examine the domain of cyber security culture, the factors essential to building and maintaining such a culture, the frameworks proposed to cultivate a security culture and the metrics suggested to measure it. Having presented the most pertinent works in the last decade, this section discusses these with the intention of providing key insights for practitioners and researchers.

# 4.1 Definitions and characteristics of cyber security culture

From analysing the studies discussing the definitions and characteristics of security culture, it becomes clear that there is no widely used definition. Although, when exploring the delineation between cyber security culture, information security culture and security culture, patterns emerge. Upon analysis of the descriptions, there are parallels in the various culture definitions identified and the broader organisational culture definitions. Nævestad et al. (2018) (P43) states that security culture can be seen simply as the security aspects of a wider organisational culture. Examining recent organisational culture literature, Oh and Han (2020) characterise organisational culture as whether employees are, "willing to participate in organisational learning activities", which corroborates Nel and Drevin's (2019) conclusion that information security culture is not only a subculture of organisational culture but should eventually become a part of organisational functions. From this perspective it is clear to recognise why so many definitions are referential to organisational culture, and the important role it plays. A takeaway point therefore is that security cultures need to be grounded in, and complementary of, wider enterprise culture efforts.

Further regarding the characterisation of cyber security culture, we find that organisational expectations are the same across industries. This often relates to the way employees behave within the workplace and is reflected within the definitions and across security culture domains. For example, Hassan and Ismail (2016) (P32) characterised information security culture as employees making information security a natural part of the way they execute daily tasks within the organisation, when pertaining to healthcare environments. Masrek et al. (2018a) (P40) conducted their work in a public sector context, stating that employees must be skilful regarding information security, alongside organisations having policies in place to safeguard protected information.

Marotta and Pearlson (2019) (P45) explore cyber security culture in the context of a large bank, characterising culture as the attitudes and values that encourage cyber-secure behaviours in employees in the organisation. Another definition (P58) describes cyber security culture as a sub-culture of organisational culture, incorporating attitudes, beliefs, values and knowledge that individuals use to interact with the organisations systems, and conduct relevant procedures, daily tasks and activities. Mainly this definition describes culture as a habitual behaviour within organisations. The similarities between these definitions across industries and domains suggests that regardless of context, organisations and their employees are expected to conduct certain safeguarding behaviours and follow the procedures in place to fulfil security culture expectations.

It is critical to mention that the emergence of cyber security culture is relatively recent; information security culture has been studied for significantly longer and dominates the review period as can be seen in Figure 4. Cyber security culture has become a more established terminology in the last decade in industry and the media (along with terms such as cyber-attack, cyber-threat, cyber-espionage) but its prevalence in academic research is limited. This was a surprising finding and raises questions about the research view (or indeed, researchers' view) of the term.

Despite this, as the cyber security culture articles increase, it is vital to understand where the differences lie between terms. Although information security culture and cyber security culture have been defined as different concepts that can overlap, these terms are often used interchangeably within the literature. Gcaza, von Solms and van Vuuren (2015) agreed that there lacks a clearly articulated definition of the cyber-security culture domain, and this needs to be addressed by removing ambiguity that surrounds the vocabulary used in regard to cyber security culture; however few studies in this review considered any differences between information and cyber security culture.

With more investigation on how the aforementioned terms relate, we may be able to gather further insight on whether cyber security culture research should garner more focus than information security culture, though the literature may already be slowly moving in this direction. With cyber security culture articles on the increase, this could eventually replace information security culture research; this may, for instance, form part of legacy cyber security culture research. The wider field of research is also pertinent to this discussion as we have seen a rise in the number of journals using the 'cyber' term, and considering interdisciplinary research (thus encompassing work on security culture), such as the Journal of Cybersecurity and the Journal of Cyber Security Technology.

## 4.2 Cyber security culture factors

To support practitioners in implementing a cyber security culture, distilling years of academic work into key factors and frameworks is essential. This paper explored the factors considered important to both building and maintaining cyber security culture. From examining the literature, the most commonly mentioned factor is top management support, and this factor appeared consistently over the ten-year period reviewed. There could be several reasons for this; without the support of management, cyber security initiatives may not appear significant to employees in comparison to their day-to-day tasks. It is with the support of management that cyber security culture can be given the attention it requires, i.e., resources will be provided and directed where needed. Therefore, management must not only guide employee attention to cyber security culture efforts but also ensure resources are managed correctly.

This need for senior support for security has challenged organisations for decades as security is often viewed as an inhibitor rather than an enabler of business. In itself, this highlights the need for a mindset change in the industry on how cyber security is perceived. We may already be witnessing a push for this change due to regulations/legislations such as the General Data Protection Regulation (GDPR) and California Consumer Privacy Act (CCPA) (considering the significant fines for organisations found to be lacking in their security) (Sirur et al., 2018; Forbes, 2020) and the growth in cyber insurance (with insurance acting as a financial backstop in cases of cyber-incidents) (Nurse et al., 2020). Both of these points have nudged cyber security towards a board-level topic that can impact the longevity of an organisation.

The research also demonstrates the need for clearer and more easily accessed policies and procedures. Greig et al. (2015) (P26) found that some employees of the retail organisation studied had never seen the company policies. Similarly, for Olivos (2012) (P10), the security policies had not been shared with employees because they were divided across multiple documents. For these studies, this demonstrated a significant issue with the extent of a security culture. It is up to both management and IT departments to ensure that employee knowledge of policy is at the required level, and individuals are exposed to up-to-date and relevant security information policies needed for their domain.

It is pertinent to note that the significance of these factors could shift due to other aspects becoming more noteworthy. Research indicates that the behavioural tendencies of employees are important as well, and potentially overlooked. Examining human aspects specifically, Van't Wout (2019) (P51) proposes that organisations should take a more customised approach to cyber security culture, and suggests that employees within the company be assessed from a more psychological perspective, looking at concepts such as personality, interests, and needs and motivations. Personality is discussed in three papers (P50; P54). All three of these articles were released in the last two years, demonstrating that taking personality into account is increasing in the literature. This is also in line with the increasing cyber security culture research in the last five years. A potential concern with this direction is regarding the privacy of employees themselves. For instance, gathering personality information could be useful in creating more customised cyber security culture programmes in organisations, but privacy concerns from employees could result in individuals not wanting to share such personal information with their employer.

Apart from personality, other factors have been discussed as important to both building and maintaining cyber security culture. Blythe et al. (2020) (P53) discuss the factor of rewards and sanctions as part of a cyber security awareness programme. A difficult question that can arise is to what extent should sanctions go for employees who fail to assimilate. And, if an employee is a high performer in their job, but fails to integrate into the enterprise's security culture (or consistently performs poorly in awareness or behaviour), what should the organisation do in response? In addition, the importance of establishing a network of champions (Alshaikh, 2020) has been identified as crucial. The main role of champions would be amplifying the security awareness messages and supporting the implementation of policies as well as the needs of employees. Overall, this approach aims to promote the development of norms within an organisation with champions acting as the role models. Work external to academia has also pointed to the importance of security champions and further supports this perspective (Infosec Institute, 2018).

Additionally, a somewhat unique factor included in four papers is national culture (P12; P35; P49; P54). The appearance of this factor, particularly in recent articles, could suggest that the cyber security culture at a national level also plays a key part of building a corporate cyber security culture. This is suggested in Gcaza, von Solms and van Vuuren (2015) where an ontology of national cyber security culture is developed, which includes concepts such as workforce education and SMMEs. Bada, Von Solms and Agrafiotis (2018) examined national cyber security awareness in African countries, and concluded that after areas of national needs for awareness campaigns are identified, different stakeholders from various sectors should become involved, and the audience therefore should include executive board members and employees of SMEs. This arguably has a trickle-down effect – when cyber security is regarded at a national level, this in turn helps establish these efforts in organisations; this can be applied to cyber security culture also. A future study could explore a similar idea by examining how national cyber security culture could be cultivated, however it is also imperative to note that building the culture of a nation would be significantly more of a challenge in comparison to an organisation.

## 4.3 Cyber security culture approaches, tools and/or frameworks

This section reflects on the approaches, tools and frameworks that have been identified in the literature. As mentioned in Section 4.1, it is difficult not to take into account the role that organisational culture plays, particularly given the studies that expand on Schein's Corporate Culture Model (P6; P10; P19) and those that suggest that security culture is a subculture of organisational culture. These studies demonstrate that there is validity in viewing cyber security culture as a subculture, however Whelan (2017) postulates that to really understand subcultures within and between organisations, there is much more research to be done in this field. For instance, only five articles in this review had explicit mention of organisational culture in their approaches, including P29 and P34.

In addition, another frequently referenced model is the STOPE framework proposed by Bakry (2004); three articles also form their foundation and expand this (P12; P20; P21). It is meaningful to note that the STOPE framework and the Corporate Culture model have features in common; AlHogail and Mirza (2014a) (P12) propose that their organisational dimension consists of four layers which are "artifacts, values, shared assumption, and knowledge", the same three levels in Schein's Corporate Culture Model and the additional fourth level from Van Niekerk and Von Solms (2010) (P6). This suggests that even across frameworks, the key aspects that build these models are often the same. For those seeking to develop security cultures therefore, this is advantageous as conflicting base models may not need to be navigated to plan an approach.

Change management is an approach that has appeared more recently in research over the last ten years, with papers such as P13, P39 and P51 which highlight the more holistic manner of cultivating and maintaining a more positive security culture. The implementation of change management frameworks is often complicated in nature because it can affect all activity in an organisation. While this approach has benefits, it must be taken into account that organisations have to be aware of the impacts of change, such as resistance to change, especially if the change is considered to disrupt the working environment (Stavros et al., 2016), as well as potential incompetence of change agents (Jalagat, 2016). Change management must be paired with knowledge from the employees and management to be able to implement and maintain change.

Arguably change management can play a role in SMEs, and how they are often overlooked in cyber security culture frameworks. Smaller organisations have a different organisational culture, down to structure, size and resources. For instance, Ling (2017) determined that SMEs find it more useful to have access to information from different departments for a more integrated and consolidated view; this potentially makes information security even more important in these smaller organisations. We should also note that research has found that only 15% of small businesses have a formal cyber incident management process (Lloyd, 2020); thus, further stressing the little emphasis that many SMEs place on security. Approaches geared towards SMEs are essential within cyber security culture research as one framework is unlikely to work for all types of organisations; a broad range of approaches and tools are crucial. These organisations often constitute a significant portion of businesses – in the UK for instance, 99.9% of businesses are SMEs (BEIS, 2020). Considering change in these enterprises, a study conducted by Stavros et al. (2016) found that employees of the SME were open to change as long as a number of factors were involved, some of which were communication, involvement and training. This further highlights the importance of organisational culture, and how despite the approach taken to building security culture, managing employees is hugely important.

We also must note the fact that many of these frameworks have not been tested in real organisations. It is essential for the conceptual models that have not been tested in organisations to conduct future studies whereby these approaches are tested in corporate companies for external validity; theory-based models do not always provide guidance on how these can be applied in an organisation. This will help with another challenging issue, which is how organisations choose which framework or tool will work best for them. Many of the tools are general and not specific to a type of firm or industry, which makes it even more difficult for companies to narrow down the best approach for their organisation. This issue highlights the importance of frameworks such as those for SMEs or for specific industries like healthcare or banking, as companies can pick a suitable approach for fostering cyber security culture in their organisations.

## 4.4 Cyber security culture metrics

When examining the metrics used to assess cyber security culture, the most apparent detail is that questionnaires and surveys are the main instrument to either assess knowledge or awareness of security policy. While these techniques cannot be undervalued (Rantos, Fysarakis and Manifavas, 2012), and despite their dominance in the literature, the use of these methods do present challenges for organisations.

Firstly, measuring knowledge can be useful in some cases but as Fertig et al. (2020) discuss in regard to security awareness, it cannot be assumed that knowledge influences behaviour. Employees may be aware of the policy, but this does not always correlate to displayed behaviour; for some organisations, we actually note that policy is instead used to identify areas of focus. This has been discussed in depth before by Bada et al. (2015), who also concluded that individuals often do not comply with defined policies or behaviours which may be expected of them. Furthermore, as we have seen in more targeted compliance-oriented research, there are a range of complex antecedents to security policy compliance (Cram, D'arcy and Proudfoot, 2019) and several theories and behaviour models to explain compliance (or the lack of it) (Moody, Siponen and Pahnila, 2018). These are ongoing deliberations and highlight the challenges around understanding and assessing actions of employees – most notably, as a platform for behaviour change.

Fertig et al. (2020) concludes that questionnaires alone to assess knowledge is not enough to measure information security awareness; it is the combination of acquiring metrics on both behaviour and knowledge that produces a sufficient measurement for security awareness, and plausibly this conclusion can be applied to cyber security culture. As an effort to include the behaviour component, for example, studies (P52; P56) have used compliance to policies and reporting to managers/compliance officers as metrics to assess employee behaviour and therefore develop a cybersecurity culture.

Arguably, studies that only use questionnaires are sufficiently measuring knowledge, but not behaviour. Nevertheless, studies such as P10, P26 and P49 that include training sessions, observation methods and roundtables, may consist of more reliable data on whether employees are truly displaying security aware behaviour in day-to-day situations. We stress that this does not mean questionnaires are not useful; Rantos, Fysarakis and Manifavas (2012) explain that it is crucial to prioritise sessions that refresh employee knowledge, because this is where the necessary security behaviour is learned. It is necessary that organisations take into account that both knowledge and behaviour need to be tested and observed where resources allow, as this will give the most accurate assessment of the state of cyber security culture.

A challenge that is faced particularly in theory-based literature is identifying metrics. Choosing a method of assessment can appear vague when metrics are not exhaustive, as seen in Reid et al. (2014) (P18) which details the General Living Systems theory. One of the processes mentioned is, "identify metrics, measure and milestones", but it does not then explain the type of metric that would be fit for measuring an information security culture system. It is important to mention that the previous step requires defining small actions that will encourage and motivate employees to create culture change, but the type of measure required for this process is unclear from the research. Despite the theory being well-developed and innovative, this demonstrates how complex it is to measure information security culture from any perspective.

Exploring whether the domains of cyber security culture are measured any differently in comparison to information security culture, there are no clear differences in terms of the metrics used. However, the differences appear in the items used in questionnaires. For example, Da Veiga (2016b) (P31) used the Information Security Culture Assessment to measure cyber security culture; the primary difference was the incorporation of elements that were related to the protection of the individual, rather than solely the protection of data and information systems. This is a useful way to demonstrate that information security culture can be incorporated into cyber security culture; as we have demonstrated earlier, the two domains possess similarities.

It is possible that the most significant issue to mention is measuring cyber security culture over time. Measuring only one point in time is a challenge that is faced with using tools such as questionnaires. These metrics provide a snapshot of an organisation's cyber security culture at a single moment, and unless they are completed at regular intervals there is a risk of not providing sufficient measurement of improvement or maintenance. One study by Da Veiga and Martins (2014) (P16) conducted assessments of an organisation's information security culture four times over eight years, which provided a longitudinal view of the company's information security culture with evidence of recommendations and improvement. When examining a corporation over time, this allows trends to be identified, and provides better evidence of whether security culture is being cultivated and maintained.

It is essential to appreciate that organisational environments (and threat landscapes) often change and as such, security cultures may need to be constantly nurtured (European Union Agency for Cybersecurity, 2018); thus, also suggesting that cyber security culture must be measured regularly. This raises a related question for practitioners and researchers about the need for more dynamic metrics and measurement tools. Organisational monitoring tools are commonplace especially for protection against both external and internal threats (e.g., Security information and event management (SEIM) software). There are, however, important considerations about the use of these tools for supporting nurturing of security cultures as opposed to detecting threats. This quickly relates to the balance between enterprise security and employee privacy.

## 4.5 Open issues in cyber security culture research

When reviewing the research, one of the challenges discussed by Marotta and Pearlson (2019) (P45) is change. This has become a widely regarded issue, evident in the increasing research on change management. Senior management teams must deal with the issue of not only fostering security culture, but discovering procedures outside of organisational frameworks that will appreciate the need and use for new policies on security (Marotta and Pearlson, 2019). Most notably, while security policies and related research has existed for decades, there are still ambiguities on how such policies are best developed (Paananen, Lapke and Siponen, 2020). Lacey (2010) (P3) takes this point further, stating that security culture research needs more focus on soft psychology ideals such as change management and education. While Section 3 has shown that training has been considered a highly essential factor to building and maintaining security culture, Pătraşcu (2019) (P48) illustrates that educating on cyber security culture is much more of a non-trivial process. Research has shown that regular training sessions often do not have the desired results; statistics show that as much as 50% of information from training sessions to improve security understanding are lost in an hour, 70% lost in 24 hours and 90% within a week (Ghafir et al., 2018).

It is also prudent to acknowledge that every organisation is different; P48 explains that corporations have their peculiarities as well as personal information and IT infrastructures, and this presents a challenge when allocating approaches for use in organisations. Each organisation needs to apply an approach or framework in a way that fits the systems and structures already in place. As Govender et al. (2020) (P55) suggest, "the key drivers to understanding, managing, supporting and changing information security culture are driven by understanding the current state of information security, investing in specific initiatives that help reduce cost while also reducing information security risk, and creating a sustainable approach to managing and improving information security culture". However, the individuality of each organisation and every employee can be viewed as an issue that needs to be addressed alongside cyber security culture research; the application of approaches is arguably as important as the approach itself. Use, testing and in-situ validation of proposed approaches is also crucial. From our review, it is apparent that few security culture approaches are evaluated (or evaluated over a notable length of time) and thus, it is challenging to understand their real value to practitioners or to the research field. Future work is desperately needed here if we are to truly address the challenges to security in organisations.

The STOPE framework by AlHogail (2015b) (P21) proposed a more complex model that allows human aspects to be considered in a more effective way. Going forward, more combinational, multidimensional frameworks that address various domains may be needed to expand on existing perspectives. There is a need to further examine personalities and individual differences such as the learning styles and individual metrics, briefly explored in Olivos (2012) (P10). Employees are different and therefore will have preferences (innate and explicit) in the training and education programmes they engage with; this also impacts any metrics used to assess behaviour. According to Fertig et al. (2020), metrics should meet the individuality of every employee, so all employees need to be considered by the metrics. Da Veiga and Martins (2015) (P25) express that there are benefits to having robust items to be applied in questionnaires across the research field to benchmark cyber security culture postures across enterprises. This difference in perspectives is noteworthy but there may be a balance that can be reached to the advantage of the entire field of research and practice. Future research should therefore explore a balanced approach between the individuality of employees and organisations and creating a standardised set of questions (which may also be open to tailoring depending on business/sector context) to assess cyber security culture.

## 4.6 Practical implications

Developing and maintaining a robust security culture has challenged businesses for decades. This has been made more complex by the abrupt move to remote working and increasing set of cyber-attacks, linked to the Covid-19 pandemic (Lallie et al., 2021). Having discussed the open research issues in Section 4.5, this section considers the implications of our work to practice and industry.

The first area in which we add value is in the up-to-date aggregation of factors and metrics core to the advancement and preservation of an organisation's security culture; also including recent discussions on cyber security culture. As has been shown, there are an increasing number of articles published on security culture, each focusing on different components (e.g., on factors, metrics or case studies). This research can help to provide a reference point for businesses which highlights primary factors witnessed over the last ten years such as top management support, accountability and trust. Moreover, we call attention to other less well-known factors including the need for a network of security culture champions or ambassadors (that can encourage and promote good security behaviour) and the underlying role that national culture plays in an enterprise's security culture. These seem underrepresented in current literature, and at least require more study. This article's findings can make a difference in security culture strategies being planned for the next decade in terms of where organisations may focus or where they may look to explore new factors.

Another important finding pertains to the fact that organisations cannot "plug and play" security culture approaches from elsewhere and expect them to work. Instead, to create a strong positive culture, businesses can begin with their current organisational culture, consider the security aspects important to them as well as the many factors identified in Section 3, and plan as appropriate. Short-term, one-off training and education campaigns have limited success in changing behaviour or engendering culture, and it is crucial that technical (e.g., types of threats) as well as softer (e.g., personality types and change management) aspects are incorporated in culture planning. Furthermore, to build or improve an organisation's culture, the development of appropriate culture metrics is essential. Surveys and questionnaires are easy to execute but have notable weaknesses (e.g., focus on quantitative data, may suffer from self-report biases, and are at a single point in time). The reality therefore is that there are no clear recommended metrics and more research is required in this domain. It would also be helpful for academics to work towards clearly defined factors (e.g., potentially looking at "trust in organisations" or "trust in employees" as factors instead of a general "trust" factor), as they can then provide a better basis for the creation of specific, more granular metrics.

To follow on from the various points above, it would also be ideal for the field if practitioners and researchers worked closer together on cyber security culture approaches, frameworks and metrics. We found an abundance of theoretical works and multiple articles which outline proposals to build or measure culture but fail to evaluate them in-situ in medium-to-long-term studies. This makes it difficult to ascertain the true value of the work and its ability to positively impact culture in organisations. If practitioners and researchers collaborate, researchers will have access to real organisations to apply, adapt, evaluate and refine their new research, and practitioners will gain access to research expertise – not always accessible – on topics from psychology and computing to sociology and organisational behaviour. This would ultimately lead to the design and development of robust set of approaches and metrics suitable for those organisations. Moreover, considering the academic impetus to publish articles and share anonymised datasets, these collaborations would also result in contributions that can advance the field for all parties involved (e.g., researchers could analyse datasets to experiment with new proposals, and companies could gain insights into how researchers and practitioners could work better together).

## 4.7 Related research reviews

As mentioned in Section 1, there have been other systematic literature reviews which have investigated the topic of security culture. It is important, therefore, to reflect on how our research contributions compare and contrast to their findings, and ultimately, what is the novelty of our research given such reviews. Three of the most recent and relevant – given their specific focus on security culture – of these articles are Glaspie & Karwowski (2017), Nasir et al. (2019a) and Sas et al. (2020).

Glaspie & Karwowski (2017) engage in a study to identify the human factors that contribute to an organisation's security culture. In this domain, they define and discuss five overarching areas, information security policy, deterrence and incentives, attitudes and involvement, training and awareness and management support. Our work has identified these areas as well, and therefore acts to further validate that research. We have also identified a range of additional factors including communication, trust, user management and national culture; all of which are increasingly pertinent to building a corporate culture of security. The topic of metrics is also not considered in Glaspie & Karwowski's research, even though there is mention of the importance of varying training according to employees' needs. Metrics are central to discussions on security culture as it is a primary way in which culture can be firstly assessed and then, improved.

In Nasir et al. (2019a), a PRIMSA-informed systematic review of the dimensions of an information security culture is presented. Their work assesses dimensions across articles and deliberates these in the context of the theories and approaches that articles ground information security culture in. There are several commonalities in dimensions (or factors, as we refer to them) across our two articles, and as to be expected considering the overlap in the date ranges (2000-2017, 2010-2020) used for article inclusion, many key references reoccur. An interesting point raised by Nasir et al. is the variation in dimensions in the literature and even disparity in the meaning of words used for what might be perceived as the same dimension/factor (e.g., trust). This is not an area that we have considered but is certainly one that deserves further research; though efforts to standardise language and terminology, even within the security field, are notoriously difficult (ENISA, 2019).

Our work also supports Nasir et al.'s earlier findings pertaining to the notable use of organisational culture as a basis for many proposed security culture models. In many ways, we view our research as complementary to Nasir et al. as we explore areas not considered in their work. For instance, we examine the changing use of terminology from security culture to information security culture and cyber security culture, and its implications for ongoing work (with 'cyber' clearly focused beyond information security alone). Moreover, we investigate the approaches/metrics (e.g., questionnaires, surveys, interviews) used to assess security culture, and identify weaknesses in current practice and thus avenues for further work (i.e., more dynamic and holistic techniques for measurement).

Sas et al. (2020) study the security culture from the perspective of measurement tools. From their search, 16 tools are identified with six then examined in detail. A key finding from their work is the attention paid to quantitative (as compared to qualitative) tools in the literature – this is comparable to the results of our study which found that surveys and questionnaires were predominately used to assess security culture. The solution to this concern is a range of metrics and tools. Sas et al. (2020) suggest a mix-methods approach. Our work, as shown in earlier sections, extends this and suggests more rounded, but also more dynamic methods of assessment. Such methods would allow a wider analysis of the organisation's security culture as well as reducing the burden on employees to actively participate – excessive active participation can also draw employees away from work which may impact job satisfaction and productivity.

Lastly, we believe there is value in the descriptive analysis that was conducted in our work which was largely absent in the other review articles covered above. From our analysis, we identified primary venues where research into security culture is published; this could serve as a guide for future authors. The examination into the use of terms, from security culture to cyber security culture, is unique to our work, as we aim to demystify such terminology, including the most noteworthy commonalities and differences. Furthermore, this study highlights the more prolific geographic centres of security culture research, with South Africa featuring prominently and a somewhat surprising lack of representation from the US, UK and Europe more broadly. Our finding would suggest the need for more research from these regions, particularly considering their role in global cyber security affairs (as superpowers, but also as countries that face significant cyber-attacks on organisations on a daily basis (CSIS, 2020)).

## 4.8 Limitations

Due to the nature of the research field, the articles were primarily dominated by information security culture (instead of security culture or cyber security culture) research. While these are all extremely relevant, looking at security culture from a wider perspective, or specifically cyber security culture, was limited and may be biased towards information security. However, within the realms of this study, information security culture still falls under cyber security culture.

Many measures were taken to ensure that all of the relevant papers within the ten-year period review were included in this study, but there exists a possibility that not all papers relating to information security culture, cyber security culture or security culture were included. Also, as mentioned, three studies were not included as they could not be accessed; therefore, it is possible that some conclusions about recent research may be missed. Considering the capture of articles across the period however, and the similarity of our findings to existing reviews, we expect that this would have had a minimal impact on our findings.

# 5. Conclusion

Cyber security culture by its nature needs to be cultivated rather than rigidly designed and this is apparent in the research. Despite the wide scope of theories and approaches covered, similar aspects and characteristics of continuously fostering culture reappear across the literature. Exploring cyber security culture (inclusive of information security culture and security culture) has allowed us to uncover the advancements as well as challenges that are currently faced in the field.

This systematic review has covered multiple research areas within cyber security culture, and from the 58 selected articles we have drawn the following conclusions:

- **Definitions and characteristics.** There are patterns between cyber and information security culture definitions. There is also a clear relationship between cyber security and organisational culture definitions. Despite these parallels, there is no universally agreed upon description. As industry moves further towards the use of the term 'cyber security culture', it will be worth monitoring how research uptake changes – if at all – as demonstrated by articles in the field. Furthermore, organisations should note that cyber security culture often focuses broader than the protection of data/information.
- **Factors.** There are multiple aspects that reappear across articles and the domains of cyber security culture. These work together; despite the frequency of top management elements, this does not cultivate good culture without other factors such as responsibility and trust, which are more personal features. Change management and national culture are more recent factors that require additional research in the future.
- **Approaches and frameworks.** Theoretical frameworks on cyber security culture dominate the research. The approaches used vary, but some struggle to be applicable across organisations sizes who do not have the same resources (e.g., SMEs versus large corporates), and industries who must manage different issues. This raises questions about the pursuit of multiple approaches which can be further tailored to a sector or company as needed.  It is also apparent that the proposal of frameworks at a conceptual / theoretical level significantly outweighs the extent of assessment in practice.  There is also a clear need for in-situ use, testing and evaluation of proposed security culture approaches and frameworks to provide real-world evidence of their efficacy.

- **Metrics.** Using questionnaires and surveys to measure knowledge of cyber security culture is mostly commonplace but does not always equate to measuring behaviour in day-to-day tasks. The only way that effective metrics will be created is through close interaction between practitioners and researchers, where practitioners can provide real-world environments for study and researchers can provide their expertise on metric design and evaluation. Such interactions will also allow medium- and long-term application and refinement of metrics, that can benefit organisations and advance the field more generally.

Overall, the overriding gap in the literature is the opportunity to consider more than one type of factor, whether it be human or technical. Case studies work well for cyber security culture research as this allows for an ethnographic environment where employees can be witnessed doing day-to-day roles and activities and provide more suitable metrics than solely tested knowledge. Therefore, next steps in research could be assessing an organisation and amalgamation of the multiple perspectives. It would also be beneficial to conduct studies on the role of psychology in employee behaviour and the protection of not only data, but cyberspace in general. Exploring whether models can be combined may be a noteworthy future study, such as Schein's corporate culture model and the STOPE framework, which share similar dimensions. It would also be beneficial to explore change management programmes in organisations. Exploring national culture should also be considered for future studies.

# Author biographies

**Betsy Uchendu** is a postgraduate student in Computer Science at the University of Kent, UK, with an undergraduate degree in Psychology. Her research interests include cyber security culture and awareness. Betsy also has marketing experience in industry covering governance, risk, compliance and technology.

**Jason R.C. Nurse** is an Associate Professor in Cyber Security in the School of Computing at the University of Kent, UK. He also holds the role of Visiting Academic at the University of Oxford, UK. He received his PhD from the University of Warwick, UK in 2010. His research interests include cyber security culture and awareness, security risk management, corporate communications and cyber security, cyber resilience, and insider threat. Jason was selected as a Rising Star for his research into cybersecurity, as a part of the UK's Engineering and Physical Sciences Research Council's Recognising Inspirational Scientists and Engineers (RISE) awards campaign. He is a professional member of the British Computing Society and an Associate Fellow of the Royal United Services Institute (RUSI).

**Maria Bada** is a Senior Research Associate at the Cambridge Cybercrime Centre of Cambridge University and a RISCS Fellow on cybercrime. She received her PhD from Panteion University of Athens, in 2013. Her research focuses on the human aspects of cybercrime and cybersecurity, as well as the effectiveness of cyber security awareness campaigns and their impact in changing online behaviour. She has collaborated with governments and International Organisations to assess national level cybersecurity capacity. She has a background in cyberpsychology, and she is a member of the British Psychological Society and the National Counselling Society.

**Steven Furnell** is a Professor of cyber security at the University of Nottingham.  He is also an Honorary Professor with Nelson Mandela University in South Africa and an Adjunct Professor with Edith Cowan University in Western Australia. His research interests include usability of security and privacy, security management and culture, and technologies for user authentication and intrusion detection. He has authored over 340 papers in refereed international journals and conference proceedings, as well as various books and book chapters. Prof. Furnell is the Chair of Technical Committee 11 (security and privacy) within the International Federation for Information Processing, and a board member of the Chartered Institute of Information Security.

# Declaration of competing interest

None.

# Appendix

The table below provides a short definition of each factor regarded as important to cyber security culture. This complements the information in Table 5. We provide definitions that capture the essence of the points made across the research articles reviewed.

| Factor | Definition |
| --- | --- |
| Top management support, leadership or involvement | The support and engagement of C-suite executives, senior managers, department managers, in creating, practicing and maintaining a security culture. |
| Security policy | A set of guidelines and processes which are defined by an organisation in relation to security. |
| Security awareness | The understanding that employees of an organisation possess regarding security generally. |
| Security training | The provision of security education materials to, and general upskilling of, employees that would make them cognisant of security threats and the organisation's policies and procedures. |
| Change management | The process that guides and supports employees towards the change necessary to develop a security culture within an organisation. |
| Compliance | The process of ensuring employees and the organisation as a whole adhere to standards and regulations on security. |
| Knowledge | Employee understanding of security hygiene practices as well as an organisation's policies and procedures. |
| Accountability and responsibility | Accountability refers to an employee owning the outcome of an action/behaviour, while responsibility refers to the employee's obligation to carry out a task that may pertain to security. |
| Security risk | This factor refers to the security threats and vulnerabilities that an organisation (and its employees) is exposed to which can lead to intentional or unintentional impacts. |
| Commitment | Employees within an organisation understand the need for security and support practises to ensure security policies are adhered to and a security culture fostered. |

| Communication | The means utilised to share information between the organisation and its employees, for instance, means through which employees find out about security policies and practices and what is expected of them. |
|---|---|
| User management | The procedures that are in place to manage and monitor employee behaviour and compliance. |
| Motivation | Incentives provided to encourage employee adherence to security policies, practices and advice. |
| Trust | Employees and the organisation need to have confidence in each other both generally and as it relates to security activities. This confidence is two-way and can relate to any activities within or about the organisation (e.g., trust in the organisation generally, trust that its policies are well considered, employees trusting their employer, etc.) |
| National culture | This relates to the norms, values, beliefs and customs of the nation or region that an organisation or employee are based in; these can influence an organisation's security culture. |
| Ethical conduct | Behaviour and decision making within an organisation follow a moral code of right and wrong. Such code can impact how people adopt or engage with a security culture (especially if that culture is perceived as unethical or harmful). |
| Regulations | The legal provisions/directives in relation to safeguarding information technology and computer systems, which organisations and their employees need to abide by. |
| Establishing a network of champions | Champions are members of an organisation who support activities in raising security awareness and act as a point of contact. Champions within different sections, departments or offices of an organisation create a network. |
| Rewards and sanctions | Utilising an approach of rewarding employee behaviour which is security compliant, or penalising non-compliance which may result in a potential compromise of the organisation. |